\documentclass[%
 reprint,
 amsmath,amssymb,
 aps,
prb,
]{revtex4-2}

\usepackage{graphicx}
\usepackage{dcolumn}
\usepackage{bm}

\begin{document}


\title{Phases of $^4$He and  H$_2$ adsorbed on doped graphene}

\author{M. C. Gordillo$^{1,2}$}
\affiliation{$^{1}$Departamento de Sistemas F\'{\i}sicos, Qu\'{\i}micos
y Naturales, Universidad Pablo de
Olavide, Carretera de Utrera km 1, E-41013 Sevilla, Spain}
\affiliation{$^2$Instituto Carlos I de Física Teórica y Computacional, Universidad de Granada, E-18071 Granada, Spain}

\author{J. Boronat$^3$}
\affiliation{$^3$Departament de F\'{\i}sica,
Universitat Polit\`ecnica de Catalunya,
Campus Nord B4-B5, 08034 Barcelona, Spain}

\date{\today}

\begin{abstract}
The influence of attractive boron impurities, embedded on a graphene sheet, on the phase diagrams of
$^4$He and H$_2$ adsorbed on top was studied using the diffusion Monte 
Carlo method. The doping of graphene was made by distributing the boron atoms 
following the same pattern found in an experimentally synthesized substrate. Our 
results show that while the different incommensurate solid phases of both adsorbates remain largely 
unchanged after doping,  the liquid/gas equations of state
are significanty different from the ones on pristine graphene.
Doping graphene produces
new translationally invariant stable phases for $^4$He,  depending on the concentration
of boron impurities,  but makes the H$_2$ ground state to remain solid.
In addition,  several new registered phases appear for both adsorbates.
 \end{abstract}

\maketitle
               
\section{Introduction}  
First and second layers of $^4$He and H$_2$ adsorbed on graphite have been 
profusely studied both
from the experimental \cite{crowell1,crowell2, 
greywall1,greywall2,vilches,wiechert,nyeki,fukuyama} and theoretical 
\cite{whitlock,corboz,PhysRevLett.81.156,
PhysRevB.59.3802,
PhysRevB.62.5228,
PhysRevLett.83.5314,
PhysRevB.65.115409,
PhysRevB.67.195411,
prl2009,prb2010,prbsecondh2,prbsecondhe4,prl2020,prb2022} 
points of view, the main motivation  
being that both are  archetypal two-dimensional (2D) many-body quantum 
systems. Both species have registered $\sqrt{3} 
\times \sqrt{3}$ solids as ground states, that change upon further loading  
into incommensurate lattices before promotion to a second layer.  In the limit 
$T = 0$, those second layers are different for both 
adsorbates: $^4$He forms a 2D liquid that solidifies upon loading,  while H$_2$ 
always remain solid.  Both second layer solids could present 
supersolidity in a narrow density range \cite{prl2020,prb2022}.
The same is true for molecular hydrogen adsorbed on a second layer of a narrow 
carbon nanotube.  On the other hand, the second layer of $^4$He on a nanotube is 
always a liquid \cite{prb2023b}.

Reducing the dimensionality of the system has been suggested as a possible way 
of getting a liquid H$_2$ phase due to a decreasing in the interaction strength.
However, two-dimensional molecular hydrogen has been always shown to be a 
solid  because of the relatively large H$_2$-H$_2$ attractive interaction 
strength,  that makes bulk H$_2$ solidify at $T>0$. Nevertheless, its low mass and 
associated large zero point motion would eventually make H$_2$ a superfluid if 
a liquid  could be supercooled up to low enough temperatures.

A possibility to frustrate the first layer solid phase would be to 
consider some kind of defects  in the substrate with respect to standard 
graphene.  For instance, one can deposit H$_2$ on a carbon glass surface 
\cite{e3}.  Diffusion Monte Carlo calculations of molecular hydrogen adsorbed on 
such environment indicate that the registered phase is substituted by a 
superglass  \cite{prb2023} with a small but still finite superfluid fraction.  
Alternatively, one can think about what would happen on novel substrates such as 
 graphyne \cite{kwon1} or  biphenylene \cite{kwon2} sheets,  where some new 
commensurate phases were found for $^4$He but that have not been 
explored as H$_2$ adsorbents yet. 

With the precedent work in mind, one realizes that an unexplored possible
way to get stable  liquid H$_2$ could be to consider its adsorption on a 
surface with impurities embedded in the substrate.  In particular,  to 
load H$_2$ on a graphene sheet in which part of the carbon atoms have been 
substituted by another species.  In this work, we will use the 
experimentally synthesized substrate of Ref. \onlinecite{boron}, that includes boron
impurities in a graphene layer. 
In order to disentangle the effect of the impurity-H$_2$ interactions from that of 
their location within the surface,  we considered both the substrate of
Ref. \onlinecite{boron} and another setup that includes a pair of boron impurities
located in opposite vertices of a graphene hexagon.  
We used a pair of impurities instead of a single one due to the 
particular structure of the boron substrate (see below). Our study was 
extended to $^4$He to compare the influence of the adsorbate-adsorbate 
interaction in the final results.   

\section{Method}

Our method is fully microscopic and starts with the Hamiltonian of the system, 
written as 
\begin{equation} \label{hamiltonian}
  H = \sum_{i=1}^N  \left[ -\frac{\hbar^2}{2m} \nabla_i^2 +
V_{{\rm ext}}(x_i,y_i,z_i) \right] + \sum_{i<j}^N V_{{\rm pair}}
(r_{ij}) \ ,
\end{equation}
where $x_i$, $y_i$, and $z_i$ are the coordinates of the each of the $N$ 
adsorbate particles with mass $m$. The potential 
$V_{{\rm ext}}(x_i,y_i,z_i)$ accounts for the interaction  
between each atom or molecule and 
all the individual atoms in the rigid graphene-with-impurities layer. 
Those potentials are of  Lennard-Jones (LJ) type, with standard parameters 
taken from Ref. \onlinecite{carlos} in the case of He-C,  and from 
Ref. \onlinecite{coleh2} for the H$_2$-C interaction.  The He-B  and H$_2$-B parameters were deduced 
using Lorentz's rules in the standard way from the C-B ones found in Ref. 
\onlinecite{parameters}. 
Those LJ parameters are
 $\epsilon_{He-B}$ = 40.2 K, $\sigma_{He-B}$ = 2.87 \AA$ $ and 
$\epsilon_{H_2-B}$ = 105.5 K, $\sigma_{H_2-B}$ = 3.10 \AA,  what implies interactions more attractive 
than for their carbon counterparts.  
We have checked the robustness of our results by varying those sets of parameters and observing the 
effects on the phase diagrams.  We found  that for any reasonable change that keeps the 
interaction more attractive than for the case of pristine graphene,   the  
phase diagrams for both of $^4$He and H$_2$ were qualitatively similar to those displayed below.

Finally, 
in Eq.  \ref{hamiltonian}, $V_{\rm{pair}}$ corresponds to the $^4$He-$^4$He and H$_2$-H$_2$ potentials.  We used the standard 
Aziz \cite{aziz} and  Silvera and Goldman \cite{silvera} models, both of 
them depending
only on the distance $r_{ij}$ between particles $i$ and $j$.  This last potential includes the $C_9$-dependent term
that effectively takes into account three-body effects. 
We used as a
simulation cell one of 51.63 $\times$ 51.12 \AA$^2$, including 1008-$N_i$ carbon 
atoms, with $N_i$ the number of boron impurities.  As indicated previously, we 
considered the case for $N_i$=2 and that of a regularly located set of boron 
atoms. To mimic the experimental compound of Ref. \onlinecite{boron}, $N_i$ 
should be 48, i.e., 5\% of the total number of atoms in the graphene-like 
layer. 

We solve the many-particle imaginary-time Schr\"odinger equation using the 
diffusion Monte Carlo (DMC) method. The DMC stochastic algorithm gives us,  
within some statistical 
noise, the exact ground-state of the $N$-particle system. In order to reduce 
the variance of the statistical estimations one uses the importance sampling 
technique. This is done by using a time-independent trial wave function which 
guides the diffusion process implied by the DMC algorithm.   
In addition, this trial wave function fixes the phase (solid or 
liquid) of
the ensemble of particles. In the present case, the trial function is written as a 
product of two terms. The first one is of Jastrow  type between the adsorbate 
particles,
\begin{equation}
\Phi_J({\bf r}_1,\ldots,{\bf r}_N) = \prod_{i<j}^{N} \exp \left[-\frac{1}{2}
\left(\frac{b}{r_{ij}} \right)^5 \right].
\label{sverlet}
\end{equation}
The values of  $b$ were the same used in previous works, i.e.,  $3.07$ \AA$ $ for the $^4$He-$^4$He case \cite{prl2009} and $3.195$ 
\AA$ $ for the H$_2$-H$_2$ pair \cite{prb2010}.
The second part incorporates the presence of C and B atoms and includes the 
possibility of localization for the solid phases, properly symmetrized, 
\begin{eqnarray}
\Phi_s({\bf r}_{1},\ldots,{\bf r}_{N})  = 
\prod_i^{N}  \prod_J^{N_S} \exp \left[ -\frac{1}{2} \left( \frac{b_{{\text
S}}}{r_{iJ}} \right)^5 \right] \nonumber \\
 \times \prod_{I=1}^{N} \left[ \sum_{i=1}^{N} \exp
\{-c [ (x_i- x_{\text{site},I})^2 + (y_i- y_{\text{site},I})^2] \} \ \right]  \nonumber \\
\times \prod_i^{N} \Psi(z_i).  \ \ \ \ \   \ 
\label{t2}
\end{eqnarray}
In Eq. (\ref{t2}), $N_S$ is the number of atoms in the substrate,  either 
carbon or boron.  The
parameters $b_S$ were taken from previous calculations on similar all-carbon substrates 
\cite{prl2009,prb2010}.   $r_{iJ}$ are the distances between a particle $i$ ($^4$He 
or H$_2$) and an atom $J$ of the substrate. On the other hand, $\Psi(z_i)$ is a 
one-body function that 
depends solely on the distance $z_i$ of every particle to the graphite 
plane \cite{prl2009,prb2010}. 

The coordinates 
$x_{\text{site,I}}$ and $y_{\text{site,I}}$
are the crystallographic positions that define the different solids (registered or incommensurate) that we are going to consider, 
and whose number is the same as the number of adsorbate particles in those structures.  
To obtain the layout of any possible commensurate solids in the 48-impurity substrate, we followed
the same procedure as in Ref. \onlinecite{prb2023} and look for local minima for 
the 
helium atoms or hydrogen molecules above the graphene-with-boron sheet. To this end, we created a 
two-dimensional grid of regularly spaced
points at a distance $z_{\text{site}}$ above the carbon-boron layer and 
calculated
$V_{\text{ext}}(x, y, z_{\text{site}})$ at such locations. $z_{\text{site}}$ 
corresponds to the maximum in the $\Psi(z_i)$ function of Eq. \ref{t2}.  
Then, we look for the place on that lattice with the minimum value of  
$V_{\text{ext}}$ and search for any other points
with the same value of the external potential located at a distance from 
the first one of, at least, $\sigma_{He-C}$ or  $\sigma_{H_2-C}$, respectively. Any set of such equal-valued potential locations will be 
the crystallographic sites of a registered phase.  Fig. \ref{fig24} displays as full squares those sites for
the lowest-density commensurate solid possible for $^4$He, while Fig. \ref{fig48} shows the 
results of an identical procedure for H$_2$.  The different sizes of the $^4$He atom and the H$_2$ molecule 
account for the number of potential minima and their different locations.  

The incommensurate solids are the standard triangular solids found for those adsorbates on graphene or graphite at 
higher densities \cite{prl2009,prb2010}.  To avoid mismatch problems between the carbon/boron substrate and
the adsorbate monolayers, we followed the procedure carried out in previous simulations including two $^4$He or H$_2$ layers on graphite \cite{prl2020,prb2022}: we took the larger piece of a triangular solid of a given density that fits in the 51.63 $\times$ 51.12 \AA$^2$ cell defined by the substrate and consider it to be at the center of a nine-cell supercell structure created by 
replicating that adsorbate simulation cell by the vectors that define its length and width.  
Obviously, we have to do the 
same with the underlying substrate using the appropriate vectors to replicate the  51.63 $\times$ 51.12 \AA$^2$ box.  Then, for each $^4$He or H$_2$ in the central cell of the triangular solid, we calculated the  adsorbate-adsorbate or adsorbate-adsorbent interactions within a given cutoff distance, irrespectively of the location of the other particle,  in the central cell or in one of the images produced by replication. 
That cutoff for the interactions, taken as half of the smallest side of the adsorbate simulation cell, should be large enough to avoid size effects. This way of dealing with the adsorbate/adsorbent of the simulation cells is completely equivalent to the approximaxion used in Refs. \cite{PhysRevLett.81.156,PhysRevB.59.3802,PhysRevLett.83.5314,PhysRevB.67.195411}, but it makes possible to consider simulation cells for adsorbate densities beyond those which fit exactly inside the periodicity of the adsorbent cell.

Finally,  the $c$ parameters in Eq. (\ref{t2}) were variationally optimized and 
found to have similar values to the ones for pure graphene.  Importantly, the form 
of Eq. \ref{t2} allows for the those solid phases to be supersolid,  since 
$^4$He atoms or H$_2$ molecules could be involved in exchanges to make them 
indistinguishable from one another \cite{claudiosuper}. 
Obviously, this is also true for translationally invariant phases for which 
$c=0$.

To avoid any influence of the initial configurations on the simulations results,  we used for the energy averages only 
the last 10$^5$ Monte Carlo configurations in a typical 1.2 10$^5$ steps long simulation run.  Each Monte Carlo step considers 300 replicas 
(walkers) to account for all the possible configurations of the system. 
The number of particles in each adsorbate simulation cell was fixed by the desired densities.  
Larger number of walkers or
longer simulation runs do not change the final results within our numerical 
accuracy.  To avoid the influence of the correlations 
between configurations in the same run we calculated all the averages using 10 independent DMC histories,
and considered only sets of positions located 100 Monte Carlo steps away. The error bars, when  given, correspond to
the averages of those 10 histories, not on averages within any single run. 

\section{Results}

\subsection{$^{\bf 4}$He}

\begin{figure}
\begin{center}
\includegraphics[width=0.8\linewidth]{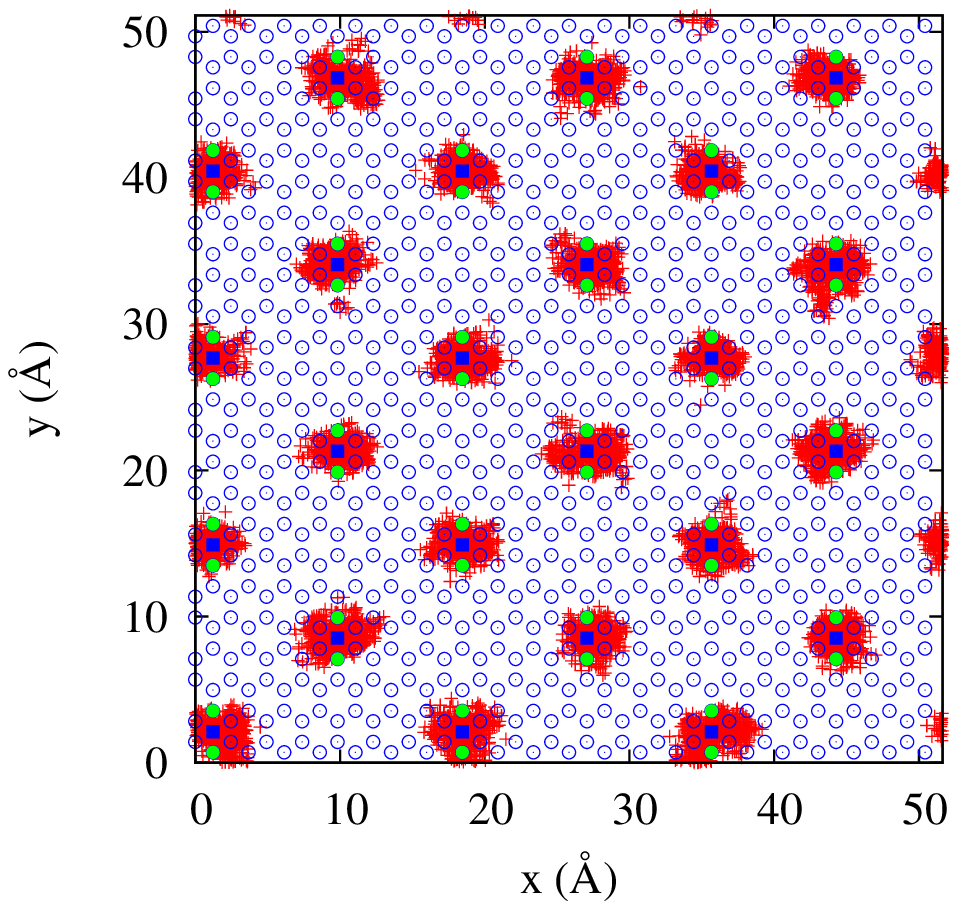}
\caption{Structure of the 48-impurities structure used as a substrate in part of 
the simulations in this work.  Open circles,  C atoms; solid circles, B atoms.  
Solid squares, positions corresponding to the 24 locations with the lowest 
$^4$He-substrate potential.  Solid smudges are the result of 
displaying 300 sets of helium coordinates represented as crosses for a 
registered phase including as many $^4$He atoms as potential minima.
}
\label{fig24}
\end{center}
\end{figure}

\begin{figure}
\begin{center}
\includegraphics[width=0.8\linewidth]{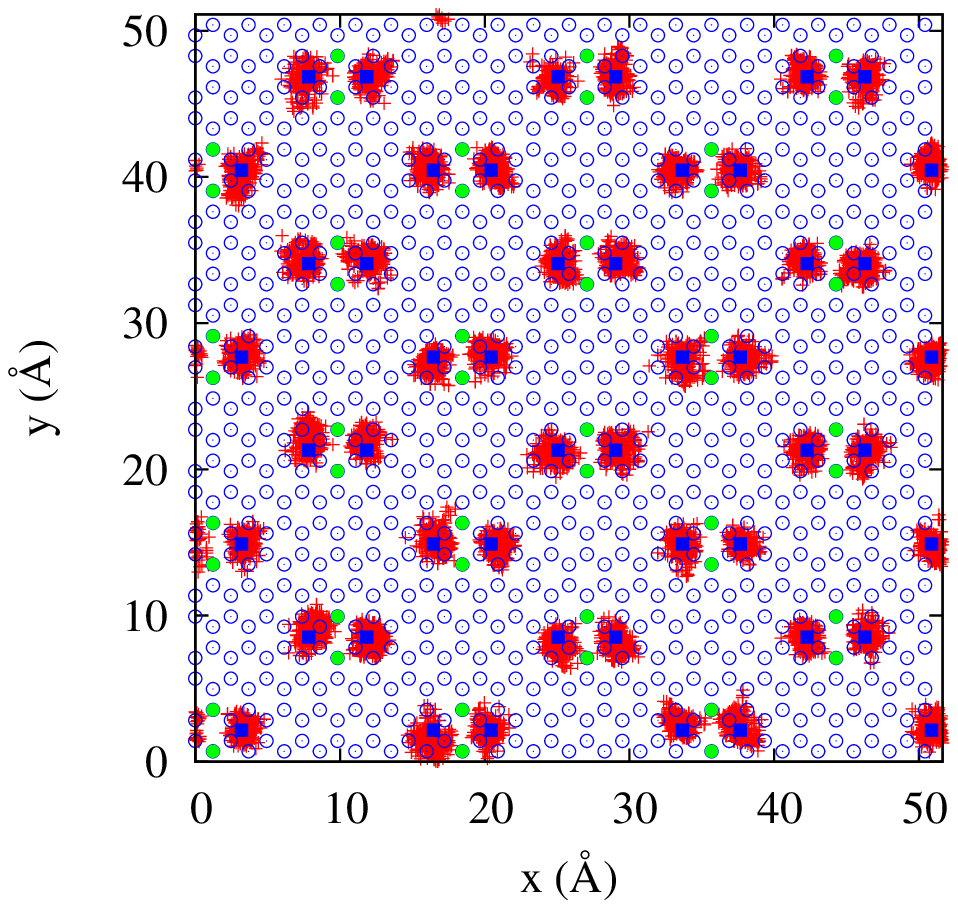}
\caption{Same as in Fig. \ref{fig24},  but for the lowest density registered H$_2$ phase.  We have 48 H$_2$-substrate potential minima, represented by solid squares.  As before,  solid smudges are the result of displaying 300 sets of hydrogen coordinates represented as crosses.
}
\label{fig48}
\end{center}
\end{figure}

The structure of the substrate corresponds to one of the possibilities experimentally synthetized in Ref. \onlinecite{boron}, 
and includes only boron impurities embedded in graphene.  Similar compounds  in which a C-C pair is substituted by a 
B-N dimer \cite{bn} can be found in the literature, but are not the object of this work. Its impurity distribution 
is depicted in Figs. \ref{fig24} and \ref{fig48}. There, the open circles stand for the carbon 
atoms and the solid ones for the boron impurities.  Those are distributed in 
pairs located in opposite vertices of the hexagons that make up the graphene 
layer.  To try to distinguish between the effect of the impurities themselves 
from that of their distribution,  we performed first  simulations including 
a single pair of impurities in the same hexagon.  After that, we 
considered all 48 boron atoms in the  51.63 $\times$ 51.12 \AA$^2$  cell shown 
in those figures. 

\begin{figure}
\begin{center}
\includegraphics[width=0.7\linewidth, angle=-90]{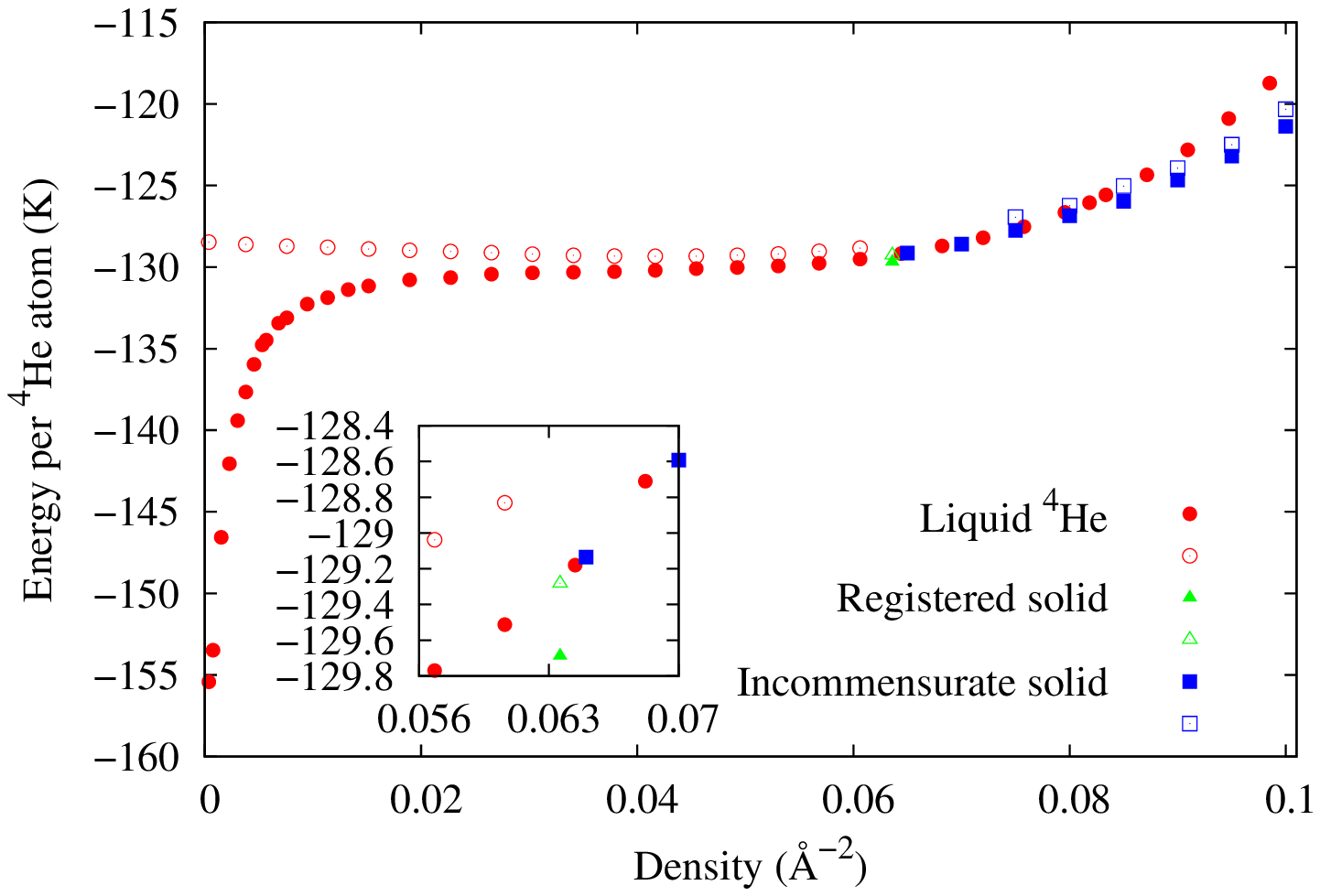}
\caption{Energy per $^4$He atom as a function of helium density for pristine graphene (open symbols) and for graphene including only two boron impurities (solid symbols). The inset shows the same data around the density corresponding to the commensurate solids. The error bars are of the size of the symbols and not shown for simplicity.
} 
\label{fig1he}
\end{center}
\end{figure}

DMC energies per $^4$He atom, as a function of the helium 
surface density $\rho_{\rm He}$, 
for the two-impurity case (full symbols) are displayed in Fig. \ref{fig1he} 
together with the equivalent results for all-carbon graphene (open symbols).  
The energies for pure graphene are identical  to the ones in  Ref.  
\onlinecite{prl2009}, obtained for smaller simulation cells. In 
all cases, the energy per particle when boron atoms are present is lower than 
for a substrate with no impurities.  This is obviously due to the larger 
$^4$He-B interaction in comparison to the $^4$He-C one.  However, we can also 
see that the difference between those two sets of energies decreases 
considerably upon helium loading.  In fact, the only obvious deviation from the 
pristine graphene case is produced at low densities and for the translationally 
invariant phase, whose energy per particle in the infinite dilution limit goes 
from -128.26 $\pm$ 0.04 K for graphene \cite{prl2009} to  -155.4 $\pm$ 0.1 K for 
the two-impurity case. 
The latter result is lower than the corresponding to a registered $\sqrt{3} 
\times \sqrt{3}$ solid on the same substrate, -129.69 $\pm$ 0.02 K. This implies 
that that commensurate structure is not longer the ground state of the helium 
monolayer. For comparison, the energy per particle of the same $\sqrt{3} 
\times \sqrt{3}$ arrangement in 
graphene is -129.282 $\pm$ 0.007 K \cite{prl2009}.

\begin{figure}
\begin{center}
\includegraphics[width=0.7\linewidth,angle=-90]{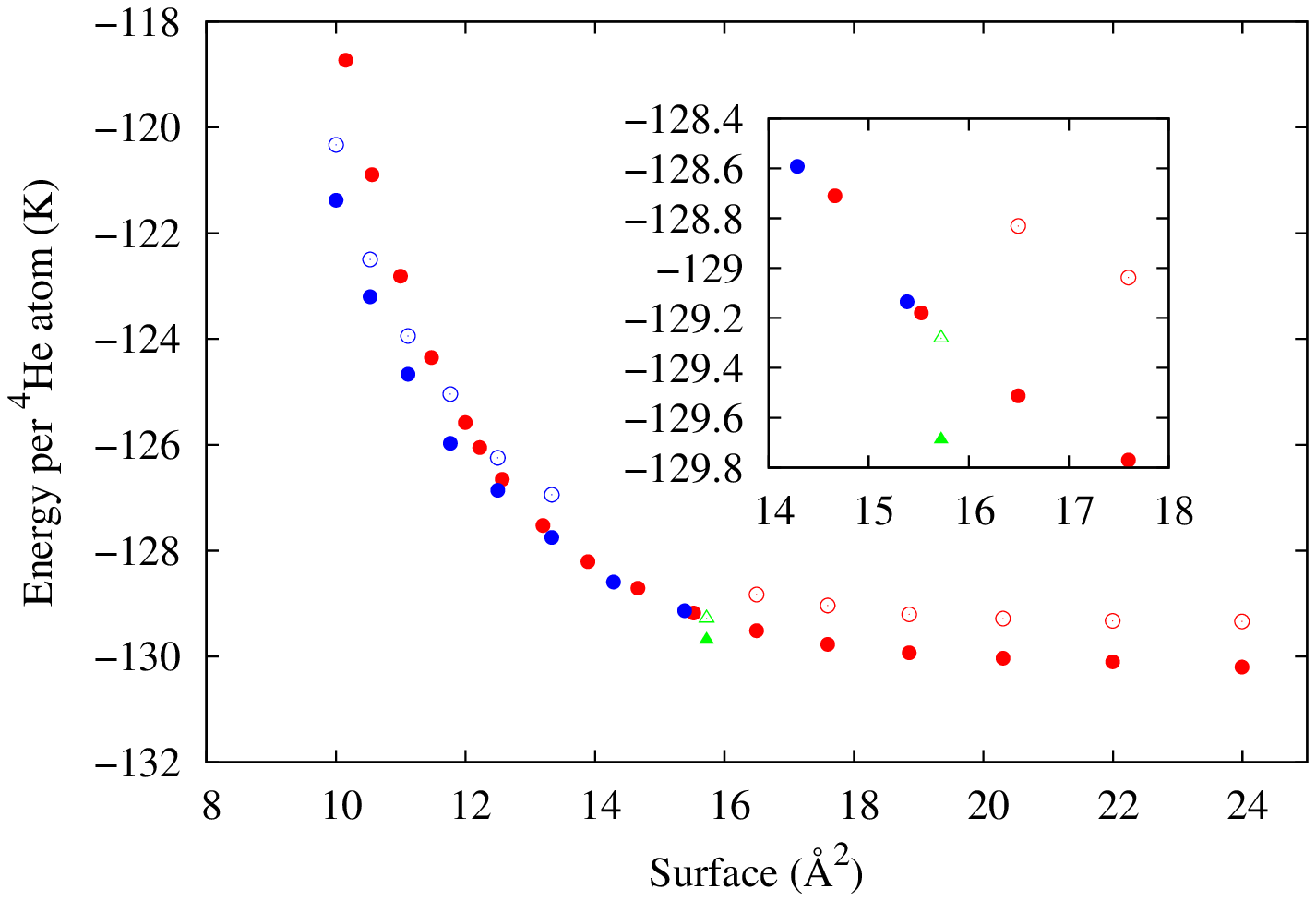}
\caption{Same as in the previous figure but displaying the energy as a function of the surface per $^4$He atom. The symbols have the same meaning as in Fig. \ref{fig1he}. Again, we show an inset around the surface per atom corresponding to the registered  $\sqrt{3} \times \sqrt{3}$ phase for both substrates.
}
\label{fig2he}
\end{center}
\end{figure}

To obtain the stability limits of the new phase diagram we need to perform 
double-tangent Maxwell constructions between the different phases.  To do so, we 
have to display the energy per particle versus the inverse of the helium density 
(or the surface per particle), in the way it is done in Fig. \ref{fig2he}. In 
this analysis, we exclude densities  $\rho_{\rm He} <$ 3.8 10$^{-3}$ \AA$^{-2}$ 
because in this regime all He atoms form clusters around the attractive impurities, i.e., do not
form an extended phase.
The Maxwell analysis shows a translationally invariant  phase for densities 0.042 $< \rho_{\rm He} 
<$0.053 \AA$^{-2}$.
 After that,  we have a first order 
phase transition to a standard $\sqrt{3} \times \sqrt{3}$ registered solid with 
density  $\rho_{\rm He}$=0.0636 \AA$^{-2}$.  That phase will be in equilibrium 
with an incommensurate arrangement of $\rho_{\rm He} >$0.070 \AA$^{-2}$ with 
energy per particle of -128.59 
$\pm$ 0.01 K stable, in principle, up to  the second layer promotion.  This is 
to be compared with a lower limit for the incommensurate structure  for graphene 
of $\rho_{\rm He}=$0.080 \AA$^{-2}$ with energy per particle of -126.6  $\pm$ 
0.2 K \cite{prl2009}.  

\begin{figure}
\begin{center}
\includegraphics[width=0.8\linewidth]{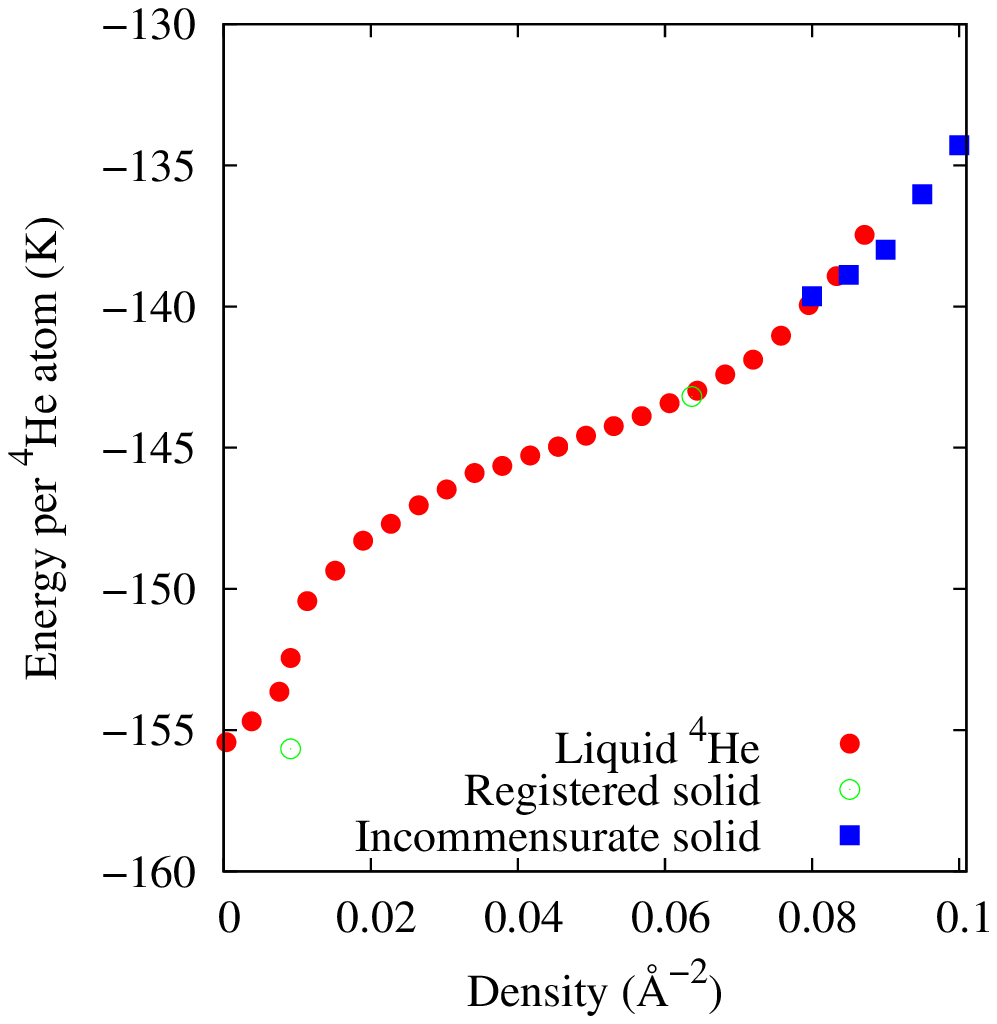}
\caption{Energy per $^4$He atom as a function of helium density for graphene including 48 boron impurities.  The open circles correspond to two different registered phases.  The solid circles are data for a translationally invariant structure, while the solid squares are the energy per atom of an incommensurate solid.  and open.  The error bars are of the size of the symbols and not shown for simplicity.
}
\label{fig3he}
\end{center}
\end{figure}

Fig. \ref{fig3he} displays the energy results for the substrate including 48 
boron atoms, 
already depicted in Fig. \ref{fig24}.  There, we can see that   the energy per 
particle is lower than the one for graphene and lower than the one for the 
two-impurity case.  In Fig. 
\ref{fig3he}, the open circles correspond to commensurate structures, while the 
solid ones show the results for the translationally invariant phase 
(circles) and the incommensurate solid (squares). 
 
\begin{figure}
\begin{center}
\includegraphics[width=0.8\linewidth]{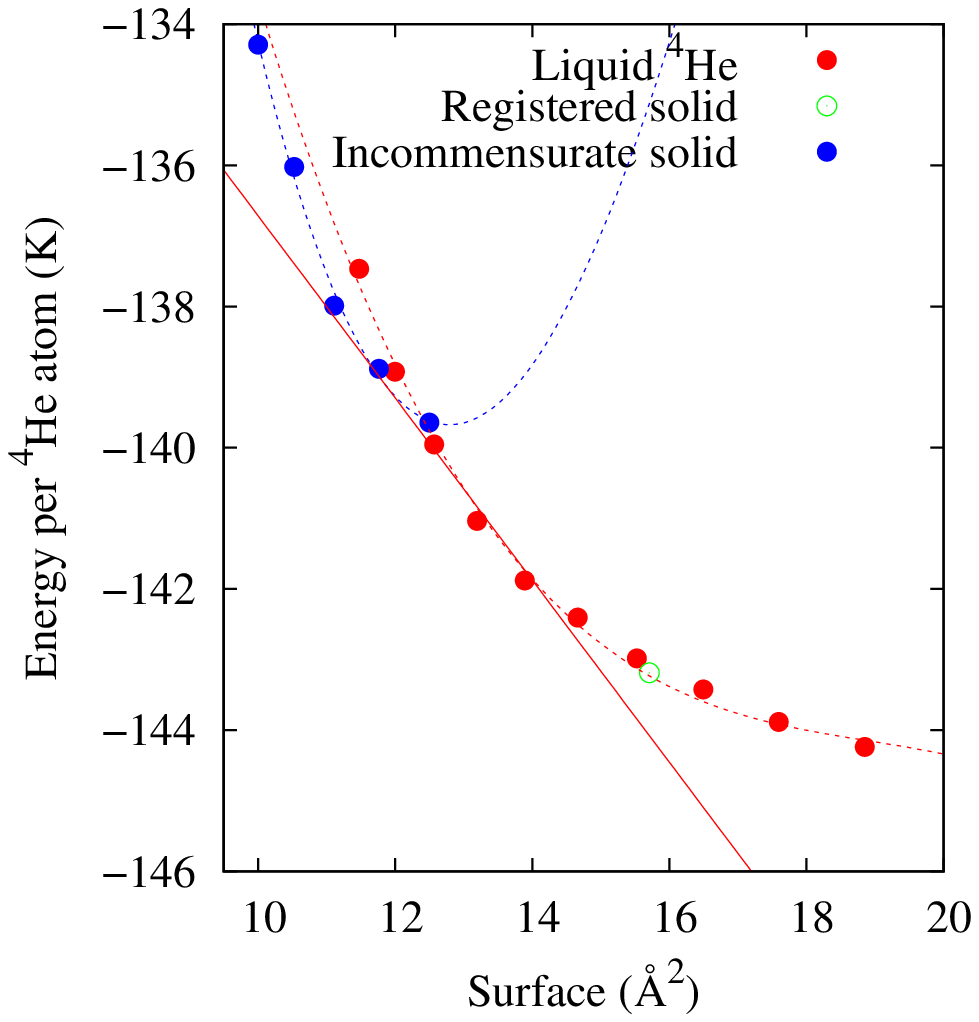}
\caption{Same as in the previous figure but displaying the energy as a function of the surface per $^4$He atom. Dotted lines are third order least-squares polynomial fits to the simulation data displayed as solid symbols.  Full line,  Maxwell construction between a liquid of density 0.072 \AA$^{-2}$ and an incommensurate solid phase of 0.085 \AA$^{-2}$. 
}
\label{fig4he}
\end{center}
\end{figure}

The registered phase with the lowest density, represented by the set of 
smudges in Fig. \ref{fig24}, is the ground state for this substrate, 
with a density of 0.009 \AA$^{-2}$ (24 $^4$He atoms in the simulation cell) and 
an energy per particle of -155.7 $\pm$ 0.1 K,  slightly lower than the 
one corresponding to the infinite dilution limit, -155.4 $\pm$ 0.1 K and equal 
to the value for  the two-impurities substrate.  This small difference could be 
assigned to the residual $^4$He-$^4$He interaction energy  of particles located 
$\sim$10 \AA$ $ apart.  The other commensurate structure corresponds to the 
$\sqrt{3} \times \sqrt{3}$ solid and, 
as it can be seen both in Fig. \ref{fig3he} and \ref{fig4he}, its energy per 
particle is slightly 
larger (-143.19 $\pm$ 0.02 K), than the corresponding to a translationally 
invariant structure (-143.24 $\pm$ 0.02 K, from the least-squares fitting line 
in Fig. \ref{fig4he}). This means that is unstable (if barely) with respect to 
that last arrangement.  No other stable registered phases were found. 

Using Figs. \ref{fig3he} and \ref{fig4he}, we can get the stability limits for 
the different 
phases on the 48-impurity substrate by means of double-tangent construction
lines, an example of which is given in Fig.  \ref{fig4he}.  The commensurate 24-atom registered solid 
is in equilibrium with a translationally invariant structure stable in the 0.038 
$< \rho_{\rm He} <$0.072 \AA$^{-2}$ range.   The energies per particle for 
those densities are -150.6 $\pm$ 0.1 K and -141.7 $\pm$ 0.1 K, respectively.  Upon further 
loading, there is a first order transition to an incommensurate triangular solid 
of density $\rho_{\rm He}$=0.085 \AA$^{-2}$ and energy per $^4$He atom of 
-138.9 $\pm$0.1 K.  This means that we have reentrant behaviour from a 
very low-density commensurate solid created by the presence of the attractive 
boron impurities to a liquid at intermediate densities. 

\subsection{H$_2$}

\begin{figure}
\begin{center}
\includegraphics[width=0.8\linewidth]{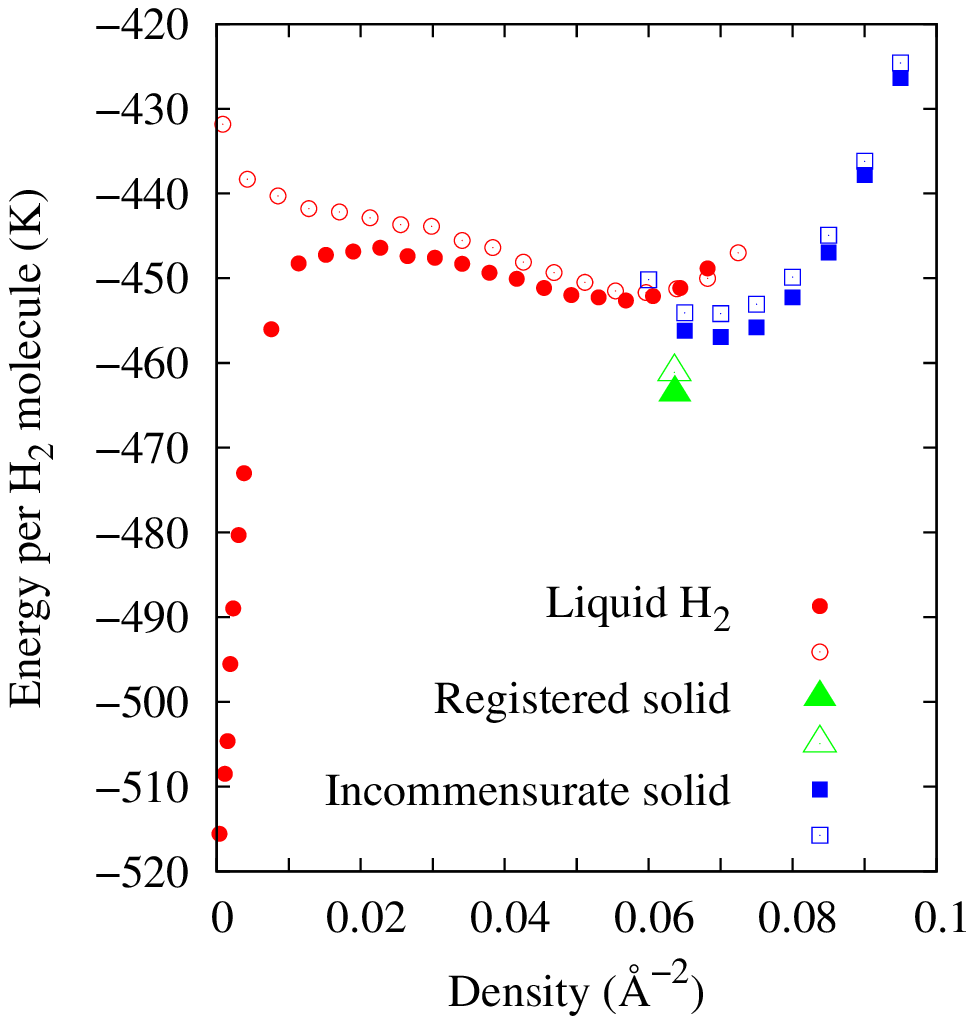}
\caption{Energy per H$_2$ molecule as a function of hydrogen density for pure graphene (open symbols) and for two-boron impurity graphene  (solid symbols).  The error bars are of the size of the symbols. 
}
\label{fig1h2}
\end{center}
\end{figure}

Fig. \ref{fig1h2} is the counterpart of Fig. \ref{fig1he} for the case of H$_2$. 
We compare the energies  for H$_2$ on  pure graphene and  with graphene 
doped with two B impurities.   As before, the energy per H$_2$ molecule 
for all values of the hydrogen density, $\rho_{\rm H_2}$, is smaller than for 
pristine graphene due to the larger H$_2$-B attraction in comparison to the 
H$_2$-C one.  Again, the infinite dilution limit when boron impurities are 
present is lower (-515.6 $\pm$ 0.1 K) than for the all-carbon substrate (-431.79 
$\pm$ 0.06 K) \cite{prb2010} and for the $\sqrt{3} \times \sqrt{3}$ commensurate 
phase, whose energy
(-463.6$\pm$0.02 K) is displayed in Fig. \ref{fig1h2} as a solid triangle.  
The open triangle in the same figure corresponds to the same structure in 
graphene (-461.12$\pm$0.01 K \cite{prb2010}). 
An analysis of the hydrogen configurations in the DMC runs indicates that for 
densities  
$\rho_{\rm H_2} <$ 0.0015\AA$^{-2}$ there is not an extended phase but a 
cluster of 3-5 molecules located close to the two-boron 
impurities The 
$\sqrt{3} \times \sqrt{3}$ registered structure is in equilibrium with an 
incommensurate triangular solid of density $\rho_{\rm H_2}$=0.075\AA$^{-2}$ and 
energy per particle  -455.84$\pm$0.02 K. That density is comparable to the one 
for the pure substrate ($\rho_{\rm H_2}$=0.077\AA$^{-2}$ \cite{prb2010}), while 
the energy per molecule in graphene is a little bit larger 
(-452.08$\pm$0.01 K \cite{prb2010}).

\begin{table}[b]
\caption{\label{tab:table1}
Densities and energies per H$_2$ molecule of the stable registered phases for the 48-impurities substrate.}
\begin{ruledtabular}
\begin{tabular}{ccc}
Number of H$_2$ molecules   & $\rho_{\rm H_2}$ (\AA$^{-2}$) & Energy (K) \\
\colrule
48 &  0.018  & -517.4$\pm$0.1    \\
96  &  0.036  &  -509.8$\pm$0.1  \\
168 ($\sqrt{3} \times \sqrt{3}$) & 0.0636 & -504.9 $\pm$0.1 \\
192 & 0.073 & -497.5$\pm$0.1 \\
\end{tabular}
\end{ruledtabular}
\end{table}

\begin{figure}
\begin{center}
\includegraphics[width=0.8\linewidth]{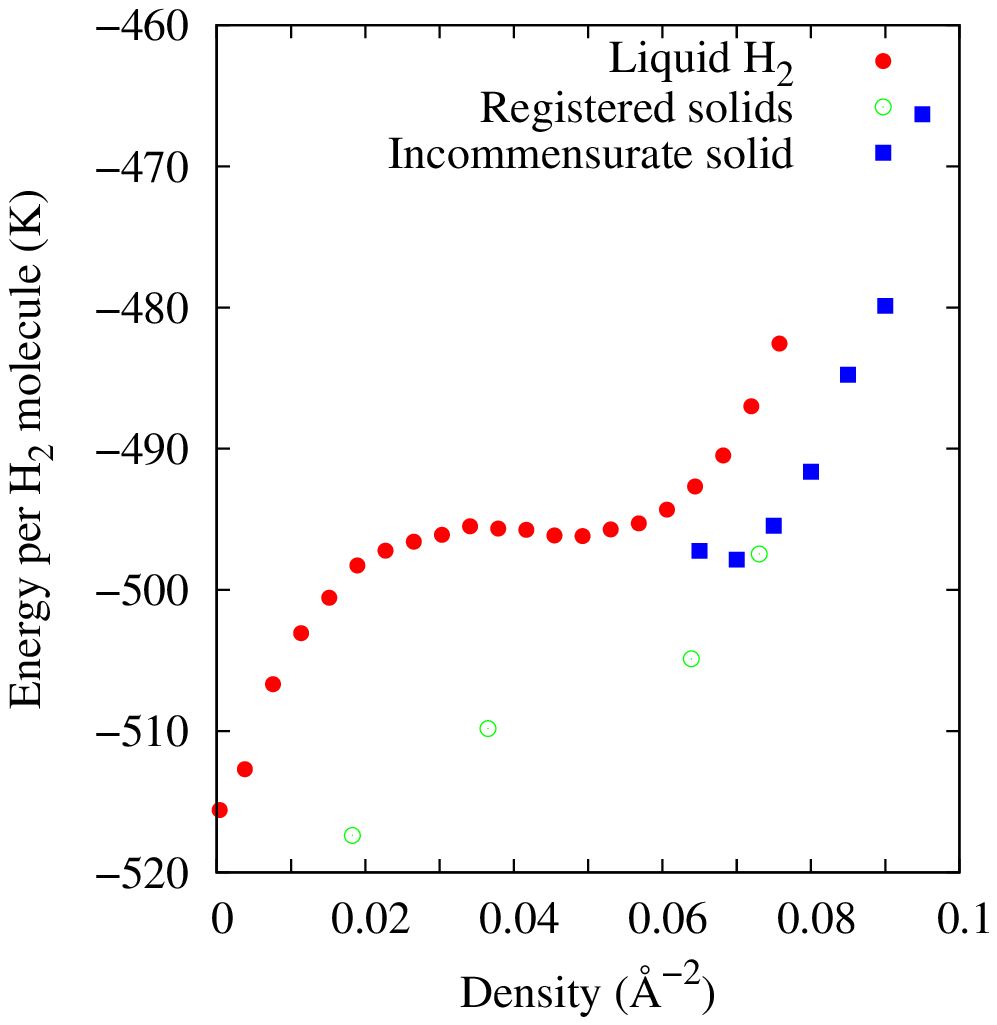}
\caption{Same as in the previous figure but for a 48-impurities substrate.  Open circles,  registered phases; solid circles,  translationally invariant structure; solid squares, incommensurate triangular solid. 
}
\label{fig2h2}
\end{center}
\end{figure}

\begin{figure}
\begin{center}
\includegraphics[width=0.8\linewidth]{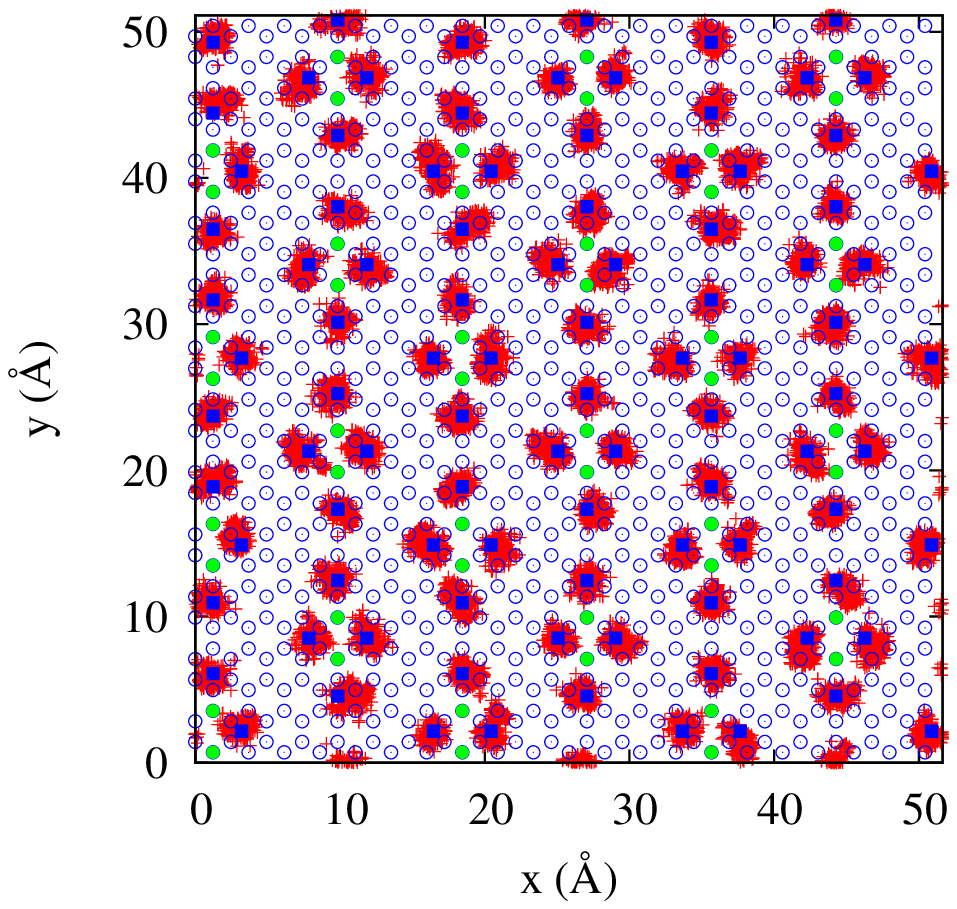}
\caption{Same as in Fig.  \ref{fig48} but for a structure with 96 potential minima.  The symbols have the same meaning as in 
that figure.
}
\label{fig96}
\end{center}
\end{figure}

\begin{figure}
\begin{center}
\includegraphics[width=0.8\linewidth]{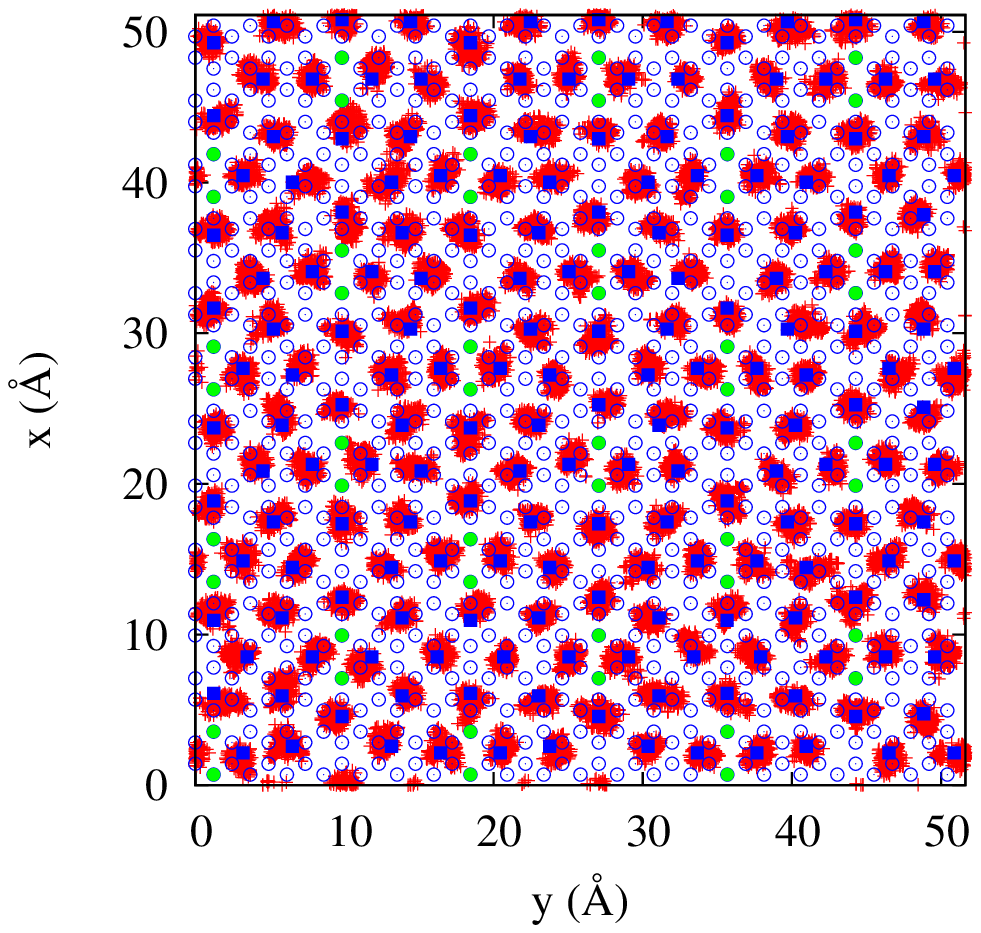}
\caption{Same as in Figs. \ref{fig48} and \ref{fig96} but for a registered solid with 192 crystallographic positions in the simulation cell. }
\label{fig192}
\end{center}
\end{figure}

The energy per particle as a function of $\rho_{\rm H_2}$ for the 48-impurity substrate is displayed in Fig. \ref{fig2h2}.  As before,  
the open circles stand for different registered phases.  The set of crystallographic positions that define those commensurate structures were obtained by the potential minima searching algorithm described above.  The only difference is that instead of having only one of such structures, we have three of them, displayed in Figs. \ref{fig48}, \ref{fig96} and \ref{fig192}. 
The remaining registered phase of density 0.0636 \AA$^{-2}$ is the standard $\sqrt{3} \times \sqrt{3}$ structure.  The different energies per particle and densities for those arrangements are listed in Table \ref{tab:table1}. 

As it is shown in Fig. \ref{fig2h2},  all the registered phases have 
energies lower than the ones for translationally invariant distributions with 
the same hydrogen density.  In particular, the most stable phase has an energy 
per H$_2$  molecule 1.8 K lower than the corresponding to the infinite dilution 
limit.  That arrangement is now the ground state and we should see upon loading 
a series of first order phase transitions between all the 
solids in Table \ref{tab:table1}.  From the last of those distributions,  we can 
draw a Maxwell construction with a common tangent to an incommensurate solid 
with $\rho_{\rm H_2}$ = 0.075\AA$^{-2}$ and energy per molecule of 
-495.5$\pm$0.4 K.  That would be the stable structure up to a second layer 
promotion.  

\section{Discussion}

Examining  the phase diagrams of both adsorbates with the 
introduction of impurities, we can see that the larger changes with respect 
to what happens in pristine graphite happen for translationally-invariant phases
at very low densities.
In those cases,  the analysis of the DMC configurations 
indicates that they are  made up of small clusters of $^4$He or H$_2$ close to a 
set of two boron impurities. This means that there is not a stable extended 
phase of the type of a conventional liquid or gas.  In addition, the energy per particle 
of those small clusters decreases drastically with respect to the infinite dilution
limit of an all-carbon substrate.   Since the influence of the boron impurities  on the
$^4$He/H$_2$ behavior decreases when the density increases, the overall
effect is of a drastic change in the equation of state of those translationally-invariant
phases.  That can be seen clearly in Figs.  \ref{fig1he}, \ref{fig3he},  \ref{fig1h2} and \ref{fig2h2}. 
In any case, those isolated 
cluster arrangements are only stable when the number of adsorbed particles and boron impurities
is very small.  When we have enough
atoms or molecules to form clusters close to neighboring pairs of boron 
atoms,
as in the case in which we have 48 boron impurities,  a commensurate low density ground 
state is produced whose nature depends on the details of the adsorbate-substrate 
potential.

Beyond those small densities,  we can see that the variations in the  $^4$He phase 
diagram depend basically on the number of boron atoms in the substrate.  
For instance,  in the two-impurities case,  the upper density limit 
is basically unchanged with respect to that 
of pristine graphene, with a stable $\sqrt{3} \times \sqrt{3}$ phase that 
transforms upon loading into a incommensurate triangular solid. 
On the other hand,  when we have 48 boron impurities, the decrease of the energy 
per $^4$He atom in the translationally invariant phase, make it 
(barely) more stable than a $\sqrt{3} \times 
\sqrt{3}$ solid of the same density.  However, the difference is small and could depend on the particular 
distribution of the boron atoms.  In any case,  we have two phases very close in energy, as in pure 
graphene \cite{prl2009}.  When the helium density increases, we end up with the same type of  
incommensurate solid structures as in an all-carbon sustrate.

For the same 48-boron substrate, the energy per particle of the  H$_2$ translationally invariant phase 
is always larger than the 
corresponding to the different registered phases.  This means that that 
arrangement is unstable, and the only difference with respect to the graphene 
case will be the appearance of other solids, i.e., a liquid H$_2$ phase cannot be
achieved by introducing attractive impurities in the substrate.  An interesting
question is the  possibility that any of those registered solids 
could be a supersolid.  The trial wave
function (\ref{t2}) allows for the possibility of exchanges between
particles,  a necessary ingredient to have supersolids.  However,
we have verified that none of the commensurate solids 
show a nonzero value for the superfluid density.
The same null result is obtained for the $\sqrt{3}
\times \sqrt{3}$ phase in the two-impurity helium case.  This means that the introduction of impurities
destroys the small
superfluid fraction found for pristine graphene \cite{claudiosuper}.  

From all the results above it is clear that, in order to get 
superfluid/supersolid behavior in H$_2$ it is not useful to
use a substrate with a
regularly disposed set of attractive impurities.  This produces only new registered
phases.  We will have to think of a situation in which no commensurate phases 
are possible,  and the equation of state of the translationally invariant structure can be changed enough to 
produce a supersolid.  Taking all of that in mind, the only successful substrate up to now
is the carbon glass already considered in Ref.  \cite{prb2023}, whose irregular structure is able to stabilize a
superglass, but not a superfluid.

\begin{acknowledgments}
We acknowledge financial support from Ministerio de Ciencia e Innovación MCIN/AEI/10.13039/501100011033
(Spain) under Grants No. PID2020-113565GB-C22 and No.PID2020-113565GB-C21 and from Junta de Andalucía group PAIDI-205. M.C.G. acknowledges funding from European regional development fund (FEDER) and Junta de Andalucía
Economy, Knowledge, Bussiness and University Consejería under ther especific goal 1.2.3 of the FEDER program Andalucía
2014-2020 “Promotion and generation of frontier knowledge and knowledge aimed at the challenges of society,
development of emerging technologies.” under Grant No. UPO-1380159, FEDER-financed percentage 80\%  and
J.B. from AGAUR-Generalitat de Catalunya Grant No. 2021-SGR-01411. We also acknowledge the use of the C3UPO
computer facilities at the Universidad Pablo de Olavide.
\end{acknowledgments}

\bibliography{irregularboro3}

\end{document}